\documentclass[a4paper, amsfonts, amssymb, amsmath, reprint, showkeys, nofootinbib, twoside,superscriptaddress]{revtex4-1}
\usepackage[english]{babel}
\usepackage[utf8]{inputenc}
\usepackage[colorinlistoftodos, color=green!40, prependcaption]{todonotes}
\usepackage{amsthm}
\usepackage{mathtools}
\usepackage{physics}
\usepackage{xcolor}
\usepackage{graphicx}
\usepackage[left=23mm,right=13mm,top=35mm,columnsep=15pt]{geometry} 
\usepackage{adjustbox}
\usepackage{placeins}
\usepackage[T1]{fontenc}
\usepackage{lipsum}
\usepackage{csquotes}
\usepackage{siunitx}

\newcommand{\silabel}{Supplemental Information}
\usepackage{hyperref} %
\usepackage{url}
\begin{document}

\title{
Extracting ice phases from liquid water: \\
why a machine-learning water model generalizes so well
}

\author{Bartomeu Monserrat}
\affiliation{Department of Materials Science and Metallurgy, University of Cambridge, 27 Charles Babbage Road, Cambridge CB3 0FS, United Kingdom}
\affiliation{Cavendish Laboratory, University of Cambridge, J.\,J.\,Thomson Avenue, Cambridge CB3 0HE, United Kingdom}

\author{Jan Gerit Brandenburg}
\affiliation{
Interdisciplinary Center for Scientific Computing,
University of Heidelberg, Im Neuenheimer Feld 205A, 69120 Heidelberg, Germany}
\affiliation{
Chief Digital Organization,
Merck KGaA, Frankfurter Str. 250, 64293 Darmstadt, Germany}

\author{Edgar A. Engel}
\affiliation{Cavendish Laboratory, University of Cambridge, J.\,J.\,Thomson Avenue, Cambridge CB3 0HE, United Kingdom}

\author{Bingqing Cheng}
\email[Correspondence email address: ]{bc509@cam.ac.uk}
\affiliation{Department of Chemistry, University of Cambridge, Lensfield Road, Cambridge, CB2 1EW, United Kingdom}
\affiliation{Cavendish Laboratory, University of Cambridge, J.\,J.\,Thomson Avenue, Cambridge CB3 0HE, United Kingdom}

\date{\today} %

\begin{abstract}
We investigate the structural similarities between liquid water
and 53 ices, including 20 known
crystalline phases.
We base such similarity comparison on the local environments that consist of atoms within a certain cutoff radius of a central atom.
We reveal that liquid water explores the local environments of the diverse ice phases,
by directly comparing the environments in these phases using general atomic descriptors,
and also by demonstrating that a machine-learning potential trained on liquid water alone can predict the densities, the lattice energies, and vibrational properties of the ices.
The finding that 
the local environments characterising the different ice phases 
are found in water sheds light on
water phase behaviors,
and rationalizes the transferability of water models  between different phases.

\end{abstract}


\maketitle
\section{Introduction} \label{sec:intro}

The number of experimentally observed and theoretically predicted phases of water seems to be ever growing~\cite{salzmann2019advances}.
The most ubiquitous phase on Earth, liquid water, has many intriguing properties, including a density maximum at $\SI{4}{\degreeCelsius}$ and ambient pressure,
volume expansion upon freezing,
unusually high surface tension, melting and boiling point~\cite{Brini2017}.
Liquid water exhibits no long-range order and its local structure is difficult to quantify and yet intricately related to its unique properties~\cite{marx1999nature,Errington2001,santra2015local,Brini2017}.
Beside the liquid, the various ice phases in the complex phase diagram of water are made from distinct local atomic environments~\cite{salzmann2019advances}, which lead to a large spread in their densities, lattice energies, and other thermodynamic as well as kinetic properties~\cite{santra2011hydrogen,salzmann2019advances}.
Apart from the direct connection with physical properties,
the local structures in water are also related to the transition paths between the phases~\cite{fitzner2019ice,del2020cubic}.

One intriguing question thus is the structural relationship between the different ice phases, and between ice and liquid water.
This is not an easy topic to investigate, however,
due to the structural complexity of the liquid~\cite{Errington2001,ohmine1999water,Ansari2019} and the large number of ice phases~\cite{salzmann2019advances}.
In this work, we exploit the state-of-the-art advances in machine learning (ML) for chemistry and materials,
in order to compare the local environments in various phases of water in a general and systematic manner.
More specifically,
we first curate a dataset consisting of 53 representative phases of ice including all the known phases (see Sec~\ref{subsec:testset}), whose densities range from 0.7 to $\SI{1.4}{\gram/\milli\liter}$.
Then we demonstrate that the local atomic environments found in liquid water cover the ones observed in \textit{all} these ice phases,
using a universal and automated framework for comparing the local similarities.
As a consequence of this inclusion, a machine-learning potential (MLP)~\cite{Cheng2019} that is only trained on liquid water accurately reproduces the ice properties including lattice energies, mass densities, and  phonon density of states.

\section{Results} \label{sec:results}

\subsection{Curated dataset of diverse water phases}

We first select representative atomistic configurations of diverse crystalline and liquid phases.
We start from 57 ice crystal structures, which include all the experimentally known ices.
These were screened from an extensive set of 15,859 hypothetical ice structures using a generalised convex hull construction (an algorithm for identifying promising experimental candidates)~\cite{anelli2018gch, engel2018ices}.
After rigorous geometry optimizations at zero pressure (see Sec.~\ref{sec:geop}), %
we eliminated three defected phases and the very high pressure phase X.
Section~\ref{subsec:testset} describes the dataset of the remaining 53 ice phases in more detail.
Note that some structures (with particular hydrogen arrangements) represent both a proton-ordered and a proton-disordered form: for example, one ice structure prototypes both ice
Ih and XI.
We consider the respective minimum potential energy configurations of the ice phases,
because they provide reasonable and reproducible approximations to the physical properties of ice, and serve as starting points for computing thermodynamic properties.

Compiling a set of representative structures for liquid water is less straightforward, since the liquid persists over a wide range of temperatures and pressures.
We consider 1,000 diverse 64-molecule snapshots of liquid water, which have previously served in training a recent MLP~\cite{Cheng2019}.
They were originally prepared using a three-step process.
Bulk liquid systems of 64 water molecules were first equilibrated at high temperatures and densities between 0.7 and $\SI{1.2}{\gram/\milli\liter}$.
The resulting (de-correlated) configurations were then quenched using a steepest decent optimization.
Finally, the 1,000 most structurally diverse structures were extracted from all the collected liquid configurations using a farthest point sampling algorithm.
There are two reasons why the 1,000 configurations are representative of liquid water.
First, they were constructed in order to cover a large part of the configurational space of possible atomic environments in liquid water.
Second, the MLP trained using these structures reproduces many properties of water very well, including the density isobar and radial distribution functions at ambient pressure~\cite{Cheng2019}, 
which means that the training set contains the necessary information for describing liquid water at ambient pressure in a data-driven manner.

\subsection{Direct comparison of the local environments} 

\begin{figure*}
    \centering
        \includegraphics[width=0.9\textwidth]{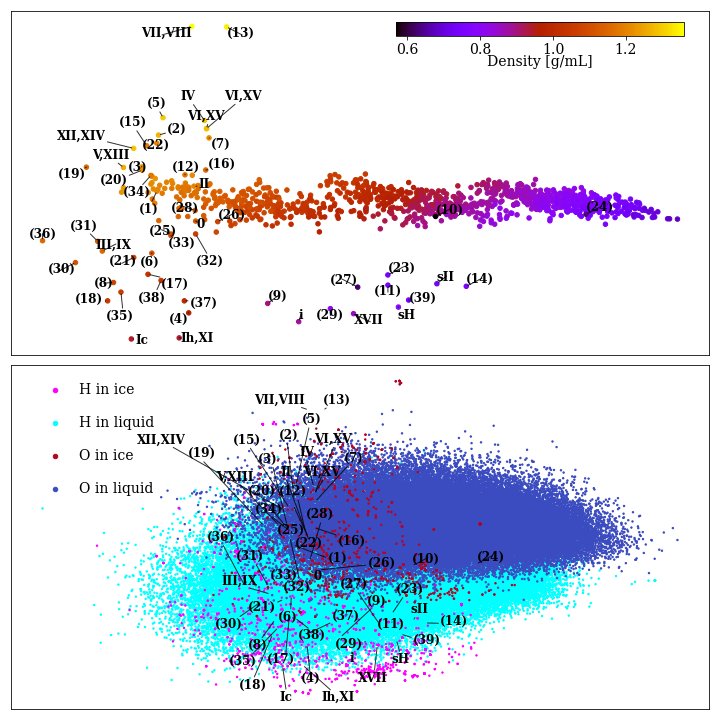}
    \caption{
    PCA maps for the 53 ice phases and the 1,000 liquid water configurations.
    The geometries of the ice phases have been optimized by HSE-3c (see Sec.~\ref{sec:geop} in Methods).
    If known, each ice structure is labelled by the name of the proton ordered and disordered phase, and otherwise by a number. 
    Upper panel: each dot indicates a structure, which is described by global descriptors.
    Lower panel: each small dot indicates the projection of a oxygen or hydrogen-centered local environment.}
    \label{fig:atomic-enviroments}
\end{figure*}

We employ the Smooth Overlap of Atomic Positions (SOAP)~\cite{bart+13prb} local descriptors to represent the atomic environments (i.e. the displacements of all the neighbors within a cutoff radius $r_\text{c}$ around the central atom).
Section~\ref{sec:representation} provides more details regarding the representations.
For each structure, we then compute its global descriptors by taking the average of the local ones of all the
atomic environments in that structure~\cite{de+16pccp}.
As the global descriptors are high-dimensional,
we use the principal component analysis (PCA) to build a two-dimensional embedding to visualise the relative difference (i.e. distances) between the structures.
Essentially, a PCA map is linear projection that best preserves the variances of the high-dimensional Cartesian distances of the dataset.
Because only linear operations are involved throughout,
the local and the global descriptors can be meaningfully projected onto the same PCA map.

We use these methodologies to analyse the 53 ice phases and the 1,000 snapshots of the liquid.
Fig.~\ref{fig:atomic-enviroments} (a) shows the PCA map of the global descriptors of all the structures:
similar structures stay close on this map,
while distinct ones are farther apart.
The horizontal principal axis is strongly correlated with density,
suggesting that density variance is a dominant feature of the dataset.
The ice phases and the liquid structures are separated on the map,
while ice structures that are commonly considered to be similar (e.g. ice Ic and Ih) stay close together.
The distinction between ices and liquids in the PCA map is to be expected 
considering the absence of long-range order in the latter.

Fig.~\ref{fig:atomic-enviroments} (b) shows the projection of local descriptors of all the atomic environments ($r_\text{c}=6$~\AA), onto the same PCA map.
The local environments in liquids and ices are similar on this map.
Furthermore, the crystalline nature of the ices leads to comparatively few distinct atomic environments,
and they are almost completely covered by the continuum realised in liquid water. 
In other words, liquid water prototypes all atomic environments pertinent to the 53 ice phases.

\subsection{Predictions of the MLP on ices}

\begin{figure*}
    \centering
    \includegraphics[width=0.45\textwidth]{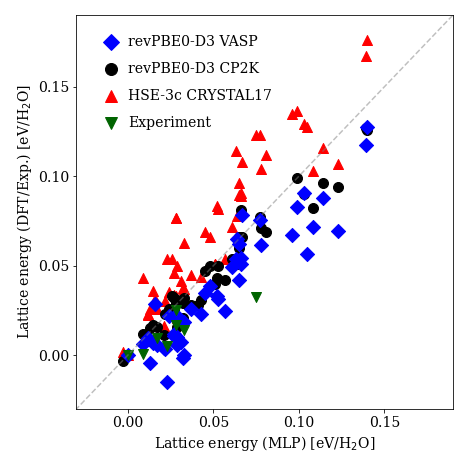}
        \includegraphics[width=0.45\textwidth]{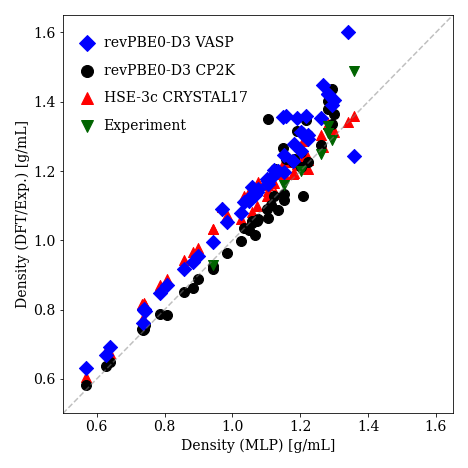}
    \caption{A comparison between the lattice and energy of the 53 ice phases computed using the MLP based on revPBE0-D3, revPBE0-D3 DFT calculations employing {\sc vasp}, revPBE0-D3 DFT calculations employing {\sc cp2k},
    and HSE-3c DFT calculations using {\sc CRYSTAL17}.
    The experimental references are taken from Ref.~\cite{Brandenburg2015}.}
    \label{fig:correlation-volumes}
\end{figure*}

The PCA maps provide a simple and general way of comparing and understanding the structural similarities,
but choice of the representations and the linear dimensionality reduction inevitably lead to information loss and distortion.
As an alternative similarity comparison, we explore how well a MLP~\cite{Cheng2019} that is only trained on reference calculations for liquid water configurations describes diverse crystalline phases. 

This MLP is based on revPBE0~\cite{revpbe,goerigk2011revpb0d3} hybrid functional DFT calculations with the semi-classical D3 dispersion correction~\cite{grimme2016vdw}.
The training set contains 1,593 configurations: the first 1,000 are classical configurations as described above,
the remaining 593 originate from path-integral molecular dynamics (PIMD) simulations at ambient conditions.
We omit those PIMD configurations in the PCA analysis above because nuclear quantum effects~\cite{markland2018nuclear} complicate the direct comparisons with classical water,
and also because those configurations had a very minor effect on the training of the MLP in our previous work~\cite{Cheng2019}.
The MLP uses an artificial neural network constructed according to the framework of Behler and Parrinello~\cite{behler2007generalized}.
The total energy of the system is expressed as the sum of the individual contributions from the atom-centered environments of radius 6~\AA{}.
Crucially, the success of the MLP hinges on the notion of ``nearsightedness'':
energy and forces associated with a central atom
are largely determined by its neighbors,
and the long-range interactions can be approximated in a mean-field manner without explicitly considering the far-away atoms.
This notion underlies many atomic and molecular force-field as well as most common MLPs~\cite{behler2016perspective}.
From this point of view, 
to capture the energetics and dynamics of a phase of water,
the key is to predict the local atom-centered contributions to the total energy the forces.
In practice, this means the training set of the MLP needs to contain the essential local atomic environments of the particular phase. 
Following this logic,
we postulate that,
if the liquid water contains all the local environments of the ice phases,
a MLP trained exclusively on snapshots of liquid water should also be able to describe the ice phases.

\begin{figure}
    \centering
    \includegraphics[width=0.45\textwidth]{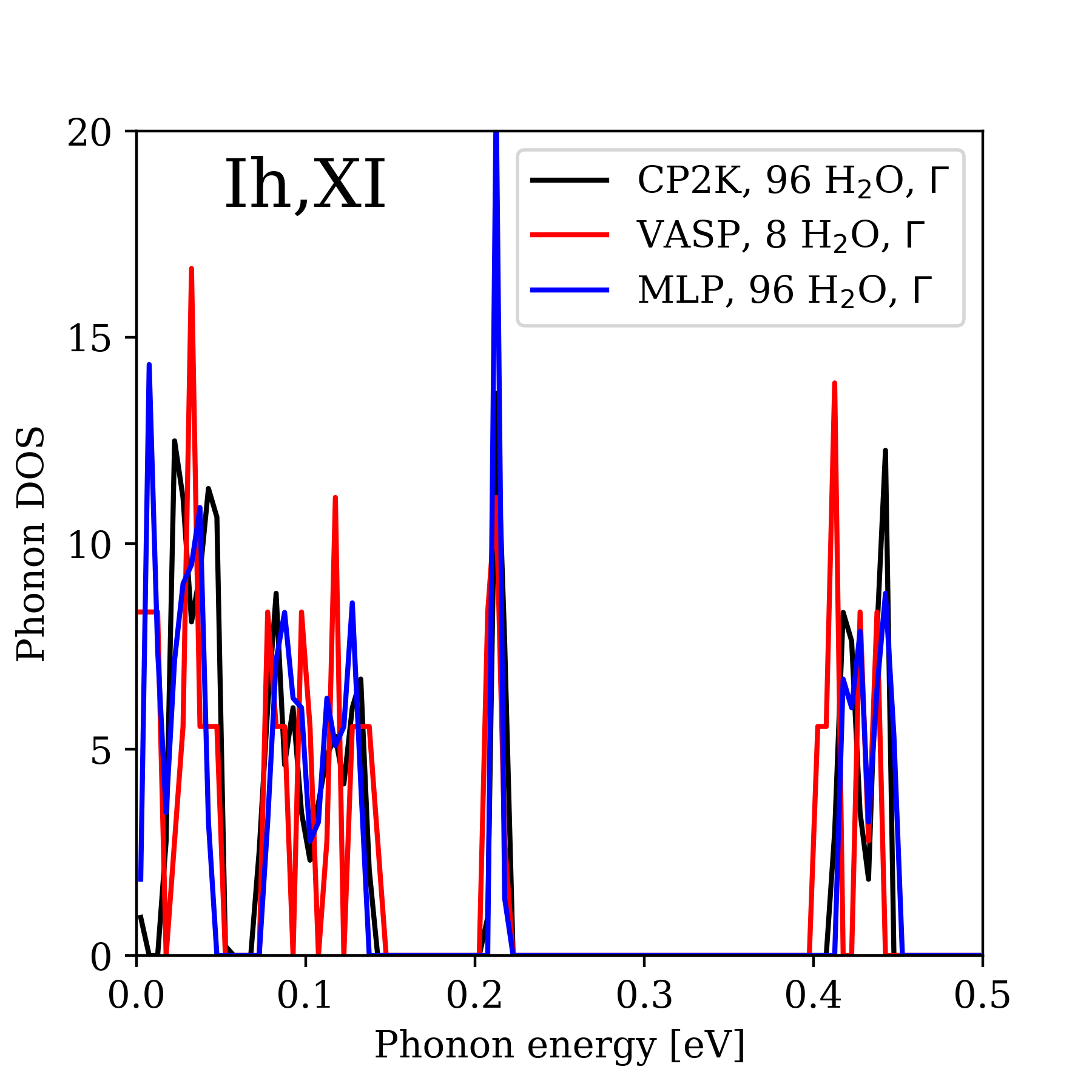}
    \caption{A comparison between the phonon density of states (DOS) for the ice Ih (XI) phase computed using revPBE0-D3 DFT calculations employing {\sc vasp} for a 8 molecule cell at the Gamma ($\Gamma$) point, revPBE0-D3 DFT calculations employing {\sc cp2k} for a 96 molecule cell at the $\Gamma$ point,
    and MLP for a 96 molecule cell at the $\Gamma$ point.}
    \label{fig:Ih-dos}
\end{figure}

To verify our hypothesis, we benchmark the performance of the MLP against reference DFT calculations and experimental results (see more details in Methods).
The DFT references comprise
(i) revPBE0-D3 using {\sc cp2k} with similar numerical settings as the calculations performed to generate the training reference of the MLP,
(ii) revPBE0-D3 using {\sc vasp} and converged numerical settings, and
(iii) the hybrid HSE-3c as implemented in {\sc Crystal17}.
Note that the multiple DFT references also provide
an estimate on the intrinsic errors in these DFT calculations due to the choices on the specific hybrid functionals, numerical settings and the use of different software packages.

\textit{Lattice energies and densities}.
In Fig.~\ref{fig:correlation-volumes} we show the comparison between the lattice energies (left panel) and the densities (right panel) of the 53 ice phases.
Note that for the lattice energies different theories or experiments have different baselines,
so for each set of calculation or measurement we use the energies the ice Ih/XI structure as a reference.
In general, we find an excellent agreement between the MLP predictions and the all four references, particularly for the densities.
In particular,
the differences between the MLP results and the \emph{ab initio} references are on par with the DFT differences introduced by the details of the first principles calculations.
For instance, the Pearson correlation coefficient $R$ between 
the MLP densities and the {\sc cp2k} values is $0.96$ and the Root Mean Square Error (RMSE) is $0.06$~g/mL,
and the corresponding metrics between {\sc cp2k} and {\sc vasp} are $0.96$ and $0.08$~g/mL.
For the lattice energies, the $R$ and RMSE between the MLP and the {\sc cp2k} predictions are $0.95$ and $9$~meV/H$_2$O, respectively, 
compared with $0.91$ and $15$~meV/H$_2$O between the {\sc cp2k} and the {\sc vasp} values.

\textit{Phonon density of states}. The curvature of the potential energy surface around a local minimum relates to the harmonic frequencies at which the atoms in a crystal vibrate. To investigate the performance of the MLP for this quantity, we have calculated the phonon frequencies for the considered ice structures using the MLP as well as for a subset of the ice structures using revPBE0-D3 DFT calculations with both {\sc vasp} and {\sc cp2k}. Figure~\ref{fig:Ih-dos} provides a detailed comparison of the phonon density of states (DOS) for a structure that represent the ice Ih phase and its proton ordered counterpart, XI, showing excellent agreement in both the low-energy region, corresponding to long-range dispersive crystal vibrations, and the high-energy region, corresponding to localized molecular vibrations. 
The small shift of low frequency phonons may be induced by the lack of long-range interactions of the MLP.
A comparison across all structures can be found in the Fig.~\ref{fig:all-dos} of Methods, and it provides remarkable agreement between the MLP and the first principles methods across all structures in the entire energy range of vibrations. 

\section{Discussion} \label{sec:discussion}

The similarities between the local environments in solid and liquid phases shed light on the
structure of liquid water.
There have been many efforts to develop a molecular understanding of water, in terms of orientational and translational order~\cite{Errington2001}, hydrogen bond networks~\cite{ohmine1999water} and spontaneously forming dendritic voids ~\cite{Ansari2019}.
Our approach of using local environments observed in ice as landmark points is a new way of interpreting liquid water as a mixture of ice structures.

On the flip side, 
the similarity also suggests 
that the liquid and ice structures are distinct in Fig.~\ref{fig:atomic-enviroments} (a)
not because of the difference in local environments,
but due to the the presence of long-range order.
The conclusion that liquid water contains all the ice environments explains why the MLP trained on liquid describes the ice phases well.
This generalization is not specific to this MLP.
Indeed, many water models, such as the coarse grained mW~\cite{molinero2009mw} model, the empirical water models SPC~\cite{berendsen1987missing} and the TIPnP series~\cite{jorgensen1983tip3p, mahoney2000tip5p}, the polarizable AMOEBA~\cite{laury2015revised},
and the MB-pol water potential~\cite{Babin2014, reddy2016accuracy} that are fitted to \emph{ab initio} reference,
qualitatively reproduce large parts of the phase diagram~\cite{yagasaki2018phasediagram, dhabal2016phasediagram}, despite having been developed primarily to simulate the liquid phase.
In particular, MB-pol correctly reproduce the properties of water from the gas to the condensed phases~\cite{reddy2016accuracy}.

In addition, the general and agnostic comparison of the local environments can be easily extended to study amorphous ice~\cite{limmer2014theory}, interfaces, and water under confinement~\cite{rossi2016nuclear}.
It is also interesting to investigate how nuclear quantum fluctuations~\cite{marx2000solvated,hbond_quantum2011,paesani2007quantum,rossi2016nuclear} influence the distribution of the atomic environments in various phases.

Furthermore, our results illustrate the immense promise of employing MLPs in materials modelling.
For instance,
the MLP used here provides an accurate description of the static and vibrational properties of ice phases at a fraction of the cost of the corresponding DFT calculations. For example, the DFT calculation for the $\Gamma$-point phonons of a 4-molecule structure takes $264$ CPU hours, and that for a 52-molecule structure takes $16,000$ CPU hours, compared with just a few minutes for both on a laptop using the MLP.
Besides,
the fact that one can only train a MLP on one liquid phase and apply the potential to other phases
evince the extend of its ``extrapolability'',
which significantly facilitates the constructions of the potentials.
It is worth noting that, the MLP for water is stable enough to run MD and PIMD for all the phases.
Last but not least,
using MLPs as tools for comparing atomic environments
offers a new approach of analyzing complex atomic systems in an agnostic and general manner.
Our analysis is, of course, 
not restricted to the chosen DFT level and can be extended to incorporate new developments in the field of \emph{ab initio} methods~\cite{wb97mv,Wang2294,riera2019low,Sharkas2020}.

\section{Conclusions} \label{sec:conclusions}

To summarize, we compare the local environments in various crystalline ice phases and liquid water, using two ML-based approaches.
We demonstrate that liquid water contains all the local atomic environments in diverse ice phases.
Our conclusion provide a new and fundamental perspective on the understanding of liquid water and ices,
and guides future efforts for modeling water.

\section{Methods} \label{sec:methods}

\subsection{SOAP representations for  atomic environments ~\label{sec:representation}}

Numerous representations of atomic environments have been developed~\cite{behl-parr07prl,bart+13prb,Faber2018,sadeghi2013metrics},
and here we use the SOAP representation~\cite{bart+13prb}.
SOAP encodes the local environment $\mathcal{X}$ around a central atom using a smooth atomic density function
\begin{equation}
    \rho_{\mathcal{X}}^{\alpha}(\mathbf{r}) = \sum_{\abs{r_i}<r_\text{c}} \exp\left(-\frac{\left[\mathbf{r}-\mathbf{r}_i\right]^2}{2\sigma^2}\right)
\end{equation}
by summing over Gaussians centred on each atom $i$ of species $\alpha$ (here hydrogen or oxygen) within a given cutoff distance $r_\text{c}$ of the central atom.
The density $\rho_{\mathcal{X}}^{\alpha}(\mathbf{r})$ is then expanded in a basis of orthonormal radial functions $g(\abs{r})$ and
spherical harmonics $Y_{lm} (\hat{r})$ as
\begin{equation}
    \rho_{\mathcal{X}}^{\alpha}(\mathbf{r}) =
    \sum_{nlm} c_{nlm}^{\alpha} g_n(\abs{r}) Y_{lm} (\hat{r}).
\end{equation}
Finally, the power spectrum is taken as
\begin{equation}
    k_{nn'l}^{\alpha} (\mathcal{X}) =
    \sqrt{\dfrac{8}{2l+1}}
    \sum_m (c_{nlm}^{\alpha})^* c_{n'lm}^{\alpha},
    \label{eqn:partial-k}
\end{equation}
which characterises $\mathcal{X}$ in a translation, permutation, and rotation invariant form~\cite{bart+13prb,de+16pccp}.
The vector $\{ k_{nn'l}^{\alpha} \}$ constructed in this way up to certain cutoffs $l_\text{max}$ and $n_\text{max}$
can then be used as the local fingerprint $\Psi (\mathcal{X})$.
We set the radius of the atomic environment to be $r_\text{c}=6$ \AA, so that it includes the second hydration shell of water molecules, and expand the SOAP descriptor up to $l_\text{max}=6$ and $n_\text{max}=6$. 
In practice, we use the DScribe Python package for constructing descriptors~\cite{dscribe},
and the ASAP Python package for the subsequent analysis~\cite{asaplib}.

\subsection{Choice of the ice configurations}
\label{subsec:testset}

\begin{figure}
    \flushleft
        \includegraphics[width=0.55\textwidth]{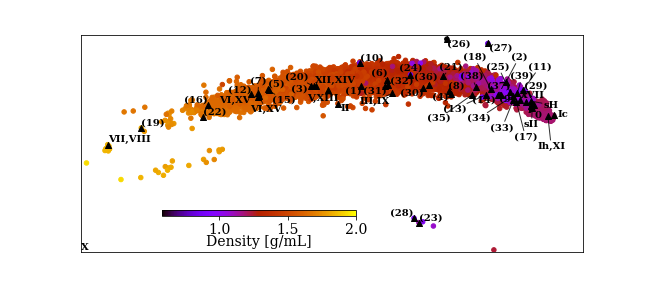}
    \caption{PCA Map for 15,859 ice phases. The phases selected for the current study, as well as the phase X (outside the map), are marked on the map.}
    \label{fig:pca-ice}
\end{figure}

The initial 57 structures are based on an extensive survey of ice~\cite{engel2018ices}
that generated 15,859 configurations,
by exploiting the isomorphism between ice,
experimentally known zeolites and theoretically-enumerated four-connected SiO$_2$ networks.
The resulting ice-like configurations were subsequently %
locally relaxed, before a generalized convex hull construction~\cite{anelli2018gch} was employed to screen for the ice structures 
that may be stable under certain thermodynamic conditions.
These include the experimentally known phases of ice except ice IV, which we add back into the selection.
Fig.~\ref{fig:pca-ice} shows the PCA map of the locations of the selected phases.
Notably, many ice phases come in pairs of a low-temperature proton-ordered form and a higher-temperature proton-disordered form: %
Ih and XI, III and IX, V and XIII, VI and XV, VII and VIII, and XII and XIV.
In this work we focus on the particular proton-ordered realizations of these phases made available with Ref.~\cite{engel2018ices}.

\subsection{Initial geometry optimization of the ice structures using HSE-3c~\label{sec:geop}}

In Ref.~\cite{engel2018ices} the ice structures were optimized at the PBE DFT level of theory using a coarse $k$-point grid and plane wave basis, trading accuracy for computational efficiency,
For this study, we have therefore performed well-converged geometry optimization for the structures,
by running a few cycles of local geometry optimizations followed by identifying and imposing crystal space group symmetries. 
These local optimizations are performed with the screened exchange hybrid functional HSE-3c~\cite{hse3c} using tight optimization thresholds as implemented in {\sc Crystal17}~\cite{crystal17, crystal17_wire}. The Brillouin zone is sampled with a $\Gamma$-centered Monkhorst-Pack grid that has been converged individually for every system to yield a lattice energy accuracy well below 1\,meV.
HSE-3c has been shown to yield excellent molecular and intermolecular geometries as well as good noncovalent interaction energies~\cite{3c-review,hse3c_hhcontact,hse3c-rev},
and in particular, 
suitable for water and ices.
The refined geometries correspond to classical 0\,K structures without external pressure and are provided in the \silabel .

\subsection{Geometry optimization using {\sc VASP}}
The geometries of all ice structures were further refined with revPBE0-D3 using the {\sc VASP} package~\cite{VASP, kresse1996software}. The equilibrium volumes critically depend on the energy cutoff, and we used a relatively high value of $1200$\,eV to obtain converged results. The $\mathbf{k}$-point sampling grids were the same as those determined for the HSE-3c calculations described above. Structures were constrained to their initial symmetries throughout the geometry optimization, and convergence was achieved with forces below $10^{-3}$\,eV/\AA\@ and stress components below $10^{-2}$\,GPa.

\subsection{Geometry optimization using {\sc cp2k}~\label{sec:cp2k}}

We computed the equilibrium densities and the lattice energies of the ice structures using the {\sc cp2k} code~\cite{lippert1999cp2k}
with the revPBE0-D3.
The computational details of the calculations are identical with Ref.~\cite{Cheng2019, marsalek2017dynamics},
although  the planewave cutoff energy was increased to $\SI{800}{\eV}$, to obtain smooth volume-energy curves.
We attached the {\sc cp2k} input file in the \silabel.
Despite a considerable amount of effort, 
the geometry optimization for 3 structures did not 
converge to reasonable values,
so these calculations were discarded.

\subsection{Phonon calculations}

\begin{figure*}
    \centering
    \includegraphics[width=0.95\textwidth]{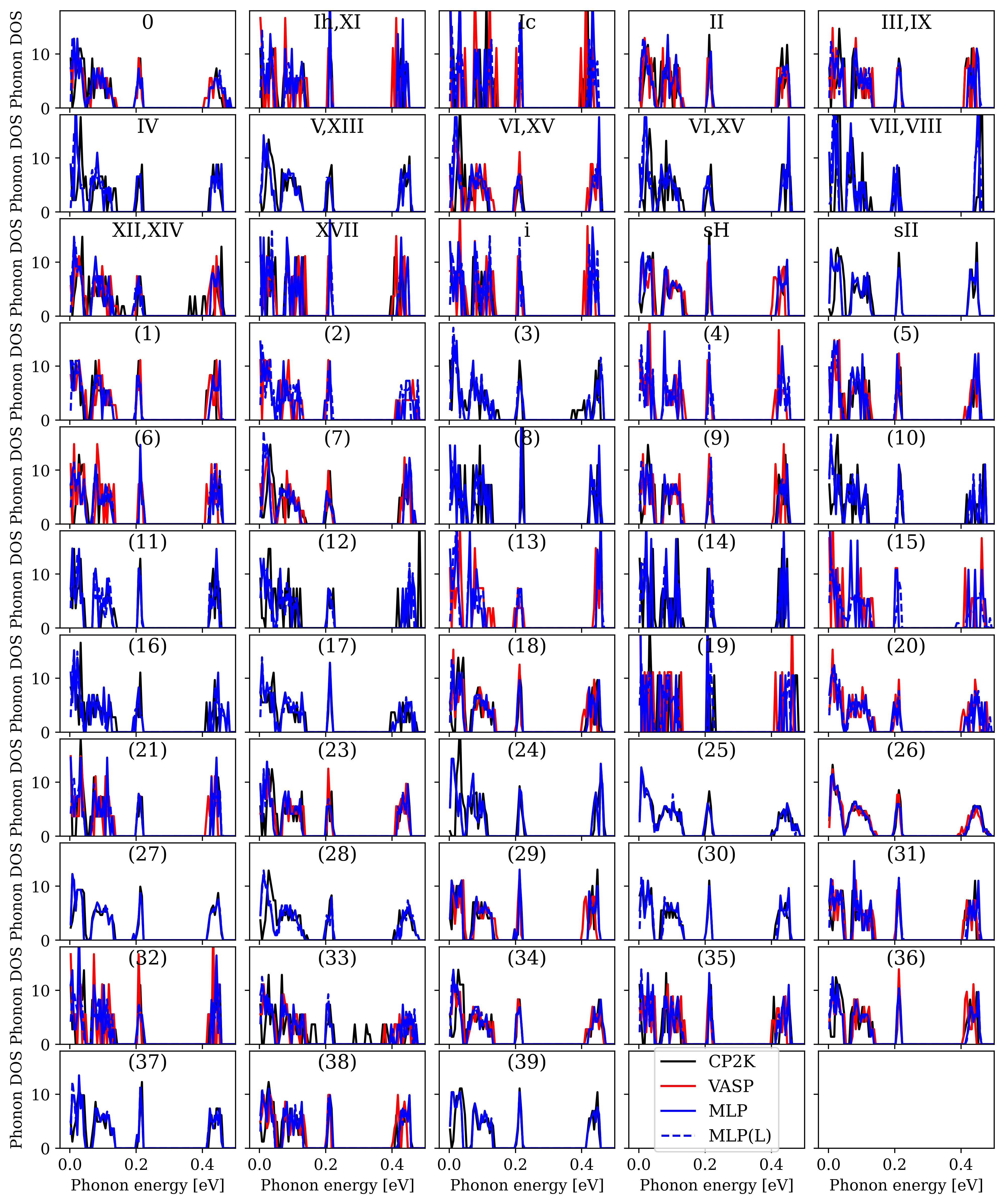}
    \caption{A comparison between the phonon spectra at the Gamma point for ice phases computed using revPBE0-D3 DFT calculations employing {\sc VASP}, revPBE0-D3 DFT calculations employing {\sc cp2k},
    MLP using the original supercell, and MLP using replicated cell (MLP(L)). For the {\sc VASP} and {\sc cp2k} calculations, only those that exhibited no imaginary phonons are shown here.
    }
    \label{fig:all-dos}
\end{figure*}

For the phonon calculations using the MLP, we first computed the Hessian matrix for the 53 geometry-optimized ice phases using finite displacements of 0.01\AA\@ of each atom from its equilibrium position along $x$, $y$ and $z$ axes.
Then the Hessian matrix was diagonalized to obtain
the phonon frequencies as the square root of the eigenvalues.
We performed those phonon calculations for the ice systems in both their original cell taken from Ref.~\cite{engel2018ices}, and in supercells of this original cell, obtained by repeating the original cell along all three crystallographic directions so that each dimension of the supercell is longer than 8~\AA. 

The phonon calculations using { \sc cp2k} follow the same approach, and the DFT settings are identical to those in Sec.~\ref{sec:cp2k}.
Presumably due to numerical issues of the specific DFT setup that we used (e.g. { \sc cp2k} only supports $\Gamma$-point sampling for hybrid functionals), a number of the ice phases contain imaginary phonons at the $\Gamma$-point, even after several rounds of geometry optimization.
We discarded the CP2K phonon DOS for these phases,
and only show the ones with real frequencies in Fig.~\ref{fig:all-dos}.

The {\sc VASP} phonon calculations were performed using the structures optimized with {\sc VASP} with the same parameters described above. We used the finite displacement method~\cite{phonon_finite_displacement} in conjunction with nondiagonal supercells~\cite{non_diagonal}, and commensurate $\mathbf{k}$-point grids were used to sample the electronic Brillouin zones of the supercells. The Hessian of a given nondiagonal supercell was calculated by displacing each atom from its equilibrium position by $0.01$\AA\@ in symmetry-inequivalent directions and calculating the force constants by finite differences. The dynamical matrix for a given $\mathbf{q}$-point grid of the vibrational Brillouin was determined by combining the results from multiple nondiagonal supercell calculations as described in Ref.~\cite{non_diagonal}. The resulting dynamical matrix was diagonalized to obtain the phonon frequencies and eigenvectors.

In all sets of phonon calculations, imaginary phonons appear in multiple structures at various $\mathbf{q}$-points in the Brillouin zone. This reflects the fact that the protons in many ice structures are disordered, and when we attempt to model them as periodic ordered structures using the unit cells from Ref.~\cite{engel2018ices}, we are artificially constraining them to a saddle point of the potential energy surface rather than to a local minimum. In some of the structures, instabilities appear even at the $\Gamma$-point, and this is caused by the symmetrization step in preparing the structures, which again can place them at a saddle point. The imaginary phonons in this case break some of the imposed symmetries to lower the energy. These problems can be resolved by replicating the original simulation cell and re-relaxing the atomic positions of the supercell to allow for the appearance of disorder that lowers the overall energy, or by re-relaxing the primitive cell without imposing symmetry in the case of $\Gamma$-point phonons. This additional step is computationally trivial for the MLP calculations, but computationally extremely costly for the hybrid functional calculations using DFT. As a consequence, it is computationally prohibitive to accurately calculate the phonon densities of states for all ice structures at the DFT level, and we only consider a subset in Fig.~\ref{fig:all-dos}.

\textbf{Data availability}
The datasets and the Python notebook for analysis are included in the Supplemental Information.
The ASAP code is available at:
\\
\url{https://github.com/BingqingCheng/ASAP}

\end{document}